
\hfuzz=4pt
\magnification=\magstep1
\font\cfont=lcircle10
\setbox4=\hbox{\lower2.5pt\hbox{\cfont\char4}}
\setbox5=\hbox{\raise1.5pt\hbox{\cfont\char5}}
\setbox0=\hbox{\lower3.2pt\vbox{ \obeylines
\copy4
\copy5}}
\setbox1=\hbox{+\kern-.5pt\copy0}
\def\ssum{\kern+2pt\copy1\kern-4pt}
\def\sdiff{{\cal SD}if\!f}
\font\titlefont=cmbx10 scaled\magstep1
\null
\vskip 2truecm
\centerline{\titlefont INFINITE DIMENSIONAL ALGEBRAS}
\centerline{\titlefont IN CHERN-SIMONS QUANTUM MECHANICS}
\vskip 2truecm
\centerline{\bf Roberto Floreanini}
\centerline{Istituto Nazionale di Fisica Nucleare, Sezione di Trieste}
\centerline{Dipartimento di Fisica Teorica, Universit\`a di Trieste}
\centerline{Strada Costiera 11, 34014 Trieste, Italy}
\bigskip
\centerline{\bf Roberto Percacci}
\centerline{International School for Advanced Studies}
\centerline{via Beirut 4, 34014 Trieste, Italy}
\centerline{and}
\centerline{Istituto Nazionale di Fisica Nucleare, Sezione di Trieste, Trieste,
Italy}
\medskip
\centerline{and}
\medskip
\centerline{\bf Ergin Sezgin}
\centerline{Center for Theoretical Physics}
\centerline{Texas A\&M University}
\centerline{College Station, TX 77843-4242}
\vskip 2truecm
\centerline{Abstract}
\smallskip
\noindent
\midinsert\narrower\narrower\noindent
We study a charged particle in an electromagnetic field without kinetic
or potential term. Although dynamically trivial, this system is interesting
because it has an infinite dimensional symmetry group.
We discuss the way in which this group behaves under quantization.
\endinsert
\vskip 2truecm
\centerline{Ref. SISSA XX/91/EP (February 1991)}
\vfill\eject

\noindent
It is a challenging task to quantize nonlinear sigma models with
an action given only by a Wess-Zumino-Witten term. We have already studied
these models from a classical point of view in [1]. Here we discuss
the quantization of the $0+1$-dimensional case.
This is the simplest version of such theories, describing a particle moving
on an $n$-dimensional manifold $N$, with action
$$S = \int dt\ \dot q^\alpha{\cal A}_\alpha(q)\  ,\eqno(1)$$
where ${\cal A}$ is a fixed background electromagnetic potential.
This model is also related to Chern-Simons theories and therefore has
come to be known as ``Chern-Simons quantum mechanics'' [2,3].
We will now recall the main features of the model, and refer the
reader to [1] for more details.

The Euler-Lagrange equations which follow from (1) are
$$\dot q^\alpha {\cal F}_{\alpha\beta} = 0\eqno(2)$$
where ${\cal F}_{\alpha\beta}=\partial_\alpha{\cal A}_\beta-
\partial_\beta{\cal A}_\alpha$ is the field strength.
These equations demand that $\dot q^\alpha$ be a null eigenvector
of ${\cal F}$. Without loss of generality, in this paper we will
consider only the case in which ${\cal F}$ is nondegenerate
({\it i.e.} it is a symplectic form).
In fact, if ${\cal F}$ was degenerate, there would be a nontrivial gauge
group ${\cal G}$ and the model would be equivalent to a particle moving
on $N/{\cal G}$ in a nondegenerate electromagnetic field.
Note that when ${\cal F}$ is nondegenerate, eq.(2) states simply that
$\dot q^\alpha=\,0$. In this case the Hamiltonian is identically zero.

The action (1) is invariant under the group $S({\cal F})$
of symplectic diffeomorphisms of $N$,
also called the canonical transformations of $N$. The Lie algebra of
$S({\cal F})$, denoted $X({\cal F})$, is the algebra of vectorfields
$v$ on $N$ such that the Lie derivative of $\cal F$ along $v$ vanishes;
equivalently, if $N$ is simply connected,
$$v^\alpha {\cal F}_{\alpha\beta}=\partial_\beta \Omega_v\ ,\eqno(3)$$
for some (globally defined) real function $\Omega_v$ on $N$
($X({\cal F})$ is called the algebra of hamiltonian vectorfields on $N$).
The Noether charge correponding to an infinitesimal symmetry
transformation $v$ is $-\Omega_v$.
The symplectic diffeomorphisms are genuine symmetries and not gauge
invariances. Thus we are in a very unusual situation in which a
finite dimensional system possesses an infinite dimensional symmetry
group.

The phase space of this theory is $N$ itself, with symplectic form
${\cal F}$ [1,4]. The Poisson bracket of two functions $f$, $g$ on $N$ is
$$\{f,g\}=({\cal F}^{-1})^{\alpha\beta}\partial_\alpha f
\,\partial_\beta g\ .\eqno(4)$$
The Lie algebra of all smooth real functions on $N$ with this bracket
will be denoted by $\Gamma$. Given any $\Omega\in\Gamma$ we can
construct a unique hamiltonian vectorfield $v$ using eq.(3).
Since functions differing by a constant give the same $v$,
$\Gamma$ is a central extension of $X({\cal F})$.
Conversely given a hamiltonian vectorfield $v$, eq.(3) defines
$\Omega_v$ only up to a constant. If it is possible to fix this
ambiguity in such a way that
$$\{\Omega_v,\Omega_w\}=\Omega_{[v,w]}\ ,\eqno(5)$$
then, as an algebra, $\Gamma={\bf R}\oplus X({\cal F})$. In this case
the center ${\bf R}$ can be dropped and the Noether charges provide a
realization of the abstract algebra $X({\cal F})$. However, as we
shall see later, this is sometimes impossible and a nonremovable central
term $c(v,w)$ may appear on the r.h.s. of eq.(5). Note that all this is
still at the classical level.

Since the dynamics of the system is trivial, the only interesting
thing to discuss are its symmetries. Note further that the classical
observables of a dynamical system are the functions on phase space;
since in our case all functions are generators of symmetries, there
follows that the discussion of the algebra of observables coincides
with the discussion of the symmetry algebra.

Now, there is a well-known theorem, originally due to van Hove,
which implies that there is no way
of quantizing this system preserving the whole classical symmetry
algebra. More precisely, there is no irreducible representation
of the observables $f\in \Gamma$ as self-adjoint operators $\hat f$
on a Hilbert space such that
$$[\hat f,\hat g]=\,i\widehat{\{f,g\}}\ .\eqno(6)$$
For a clear exposition, see [5]. This impossibility of preserving a classical
symmetry algebra at the quantum level is reminiscent of anomalies in gauge
theories. However, the situation here is quite different. In gauge
theories the anomalies manifest themselves as non conservation of
certain charges. On the other hand in our model all Noether charges
are conserved because of the vanishing Hamiltonian. Instead, the
algebra obeyed by these charges is modified.

In what follows we will consider various special cases for $N$ and
discuss the fate of the symmetry group when the model is quantized.

\medskip
$N={\bf R}^2$
\smallskip
\noindent
We take ${\cal A}_\alpha=-{1\over2}\varepsilon_{\alpha\beta}q^\beta$, with
$\alpha,\beta=1,2$ and $\varepsilon_{12}=1$.
Then ${\cal F}_{\alpha\beta}=\varepsilon_{\alpha\beta}$ is the
euclidean invariant volume form on ${\bf R}^2$ and $S({\cal F})$
is the group $\sdiff {\bf R}^2$ of volume-preserving
diffeomorphisms of ${\bf R}^2$.
A constant factor $g$ in front of ${\cal F}$ can be absorbed by a
rescaling of the coordinates. Thus the theory is independent of the
strength of the magnetic field.

The maximal finite dimensional subgroup of $\sdiff {\bf R}^2$
is the semidirect product of the group ${\bf R}\times{\bf R}$ of
translations (with generators $v^{(1)}=\partial_1$,
$v^{(2)}=\partial_2$)
and the group $SL(2,{\bf R})\sim Sp(1,{\bf R})$ (with generators
$v^{(0)}=q^2\partial_2-q^1\partial_1$, $v^{(+)}=q^2\partial_1$ and
$v^{(-)}=q^1\partial_2$). The nonvanishing Lie brackets of these
vectorfields are
$[v^{(+)},v^{(-)}]=v^{(0)}$,
$[v^{(0)},v^{(\pm)}]=\pm 2 v^{(\pm)}$,
$[v^{(1)},v^{(0)}]=-v^{(1)}$,
$[v^{(2)},v^{(0)}]=v^{(2)}$,
$[v^{(1)},v^{(-)}]=v^{(2)}$,
$[v^{(2)},v^{(+)}]=v^{(1)}$.
Note that $v^{(1)}$, $v^{(2)}$
and $v^{(-)}-v^{(+)}$ generate the Euclidean group.

Already at the classical level, the Noether charges do not give a
representation of the symmetry algebra $X({\cal F})$, but rather
of its central extension $\Gamma$.
A basis for $\Gamma$ is given by the monomials
$f^{(m,n)}=(q^1)^m(q^2)^n$, with $m,n=0,1,2\ldots$ .
The functions which generate the finite dimensional subgroup are
$\Omega_{(1)}=f^{(0,1)}$, $\Omega_{(2)}=-f^{(1,0)}$,
$\Omega_{(0)}=-f^{(1,1)}$, $\Omega_{(+)}={1\over2}f^{(0,2)}$,
$\Omega_{(-)}=-{1\over2}f^{(2,0)}$.
The Poisson algebra of these generators is the same as the Lie bracket
algebra of the corresponding vectorfields, except for
$\{\Omega_{(1)},\Omega_{(2)}\}=-1$.
Clearly the center cannot be eliminated by redefining the generators.
Therefore in this model the generators of translations fail to commute
already at the classical level and the algebra of translations is
enlarged to the Heisenberg algebra $h(1)$.

The system can be quantized in the Schr\"odinger picture replacing
$q^2$ and $q^1$ by the operators
$\hat q^2=q$ and $\hat q^1=-i{\partial\over\partial q}$
acting on the Hilbert space of square integrable functions of $q$.
The quantization of all functions in $\Gamma$ can be achieved by
replacing each monomial $f^{(m,n)}$ by the symmetrically ordered operator
$\hat f^{(m,n)}=S^{(m,n)}_{\alpha_1,\ldots,\alpha_{m+n}}
\hat q^{\alpha_1}\ldots \hat q^{\alpha_{m+n}}$,
where $S^{(m,n)}$ is the totally symmetric tensor with components
$$S^{(m,n)}_{\alpha_1,\ldots,\alpha_{m+n}}=
\cases{{m!n!/(m+n)!}\ & if $\alpha_i=1$ for $m$ values of $i$;\cr
0\ & otherwise.\cr}\eqno(7)$$
The algebra of these operators closes, but is different from the
classical symmetry algebra. We begin by observing that
the maximal finite dimensional subalgebra is not modified by
quantization; in particular, the operators $\hat \Omega^{(0)}$ and
$\hat\Omega^{(\pm)}$ give a representation of the algebra
$sl(2,{\bf R})$ with Casimir operator
${1\over2}(\hat\Omega_{(+)}\hat\Omega_{(-)}+\hat\Omega_{(-)}\hat\Omega_{(+)})
+{1\over4}(\hat\Omega_{(0)})^2={3\over16}$.
This is known as the metaplectic representation [6].
The complete quantum symmetry algebra is the universal enveloping
algebra of $h(1)\ssum sl(2,{\bf R})$.
The realization of this algebra on the Hilbert space has a nontrivial
kernel generated by the elements of the form
$$\hat\Omega_{(0)}-{1\over2}
(\hat\Omega_{(1)}\hat\Omega_{(2)}+\hat\Omega_{(2)}\hat\Omega_{(1)})
\quad , \quad
\hat\Omega_{(+)}-{1\over2}(\hat\Omega_{(1)})^2
\quad , \quad
\hat\Omega_{(-)}+{1\over2}(\hat\Omega_{(2)})^2 \ .\eqno(8)$$
The algebra which is realized faithfully is obtained by factoring
out this kernel. It is isomorphic to the universal enveloping algebra
${\cal U}(h(1))$.

We note that any ordering prescription would give rise to a closed algebra
of quantum operators. All these algebras are isomorphic.
In fact, a choice of ordering prescription is equivalent to a choice
of basis in the universal enveloping algebra.

Another possibility is to introduce a ${\bf Z}_2$ grading in the space
of the quantum operators $\hat f^{(m,n)}$: we call them fermionic or
bosonic depending on whether $m+n$ is odd or even. Then, introducing
a corresponding graded bracket, the operators $\hat f^{(m,n)}$ with
$m+n=1,2$, generate the noncompact version of the superalgebra
$osp(1,2)$. Note that in this approach it is not necessary to
introduce the constant multiples of unity to close the algebra.
This representation of $osp(1,2)$ is also known as the
metaplectic representation [7]. With this graded bracket the entire
algebra generated by the operators $\hat f^{(m,n)}$, with $m+n>0$, is the
universal enveloping algebra of $osp(1,2)$. Note that this is also
known as the higher spin algebra $shs(2)$ [8].
Again, there is a nontrivial kernel given by (8).

Finally, we observe that all that has been said so far for the case
$N={\bf R}^2$ can be generalized in a straightforward manner to the
case $N={\bf R}^{2n}$. We choose ${\cal F}_{2i-1,2i}=
-{\cal F}_{2i,2i-1}=1$, for $i=1,\ldots,n$. Then, the quantum symmetry
algebra is the universal enveloping algebra of
$h(n)\ssum sp(n,{\bf R})$ if only commutators are used, and of
$osp(1,2n)$ if the graded brackets are used.
\goodbreak

\medskip
$N=S^2=SO(3)/O(2)$
\smallskip
\noindent
In this case we take ${\cal F}$ to be the field strength of a monopole
with magnetic charge $g$. Unlike the previous case, here it is not
possible to eliminate the constant $g$ from the discussion
by redefining coordinates.
In polar coordinates $\theta$ and $\varphi$,
${\cal A}_\theta=\, 0$, ${\cal A}_\varphi=-g(1+\cos\theta)$, and
${\cal F}_{\theta\varphi}=g\sin\theta$. The symmetry group
$S({\cal F})$ is now the group $\sdiff S^2$ of volume-preserving
diffeomorphisms of the sphere. Its maximal finite dimensional subgroup
is $SO(3)$.

The Noether charges can be uniquely defined by requiring that their
integral over $S^2$ be zero. In this way no center arises [1].
This gives a realization of
the Lie algebra of $\sdiff S^2$ as the Poisson
algebra of functions on $S^2$ with vanishing integral.
A suitable basis for these functions are the spherical
harmonics $Y_l^m$ with $l=1,2\ldots$.
The structure constants of the Lie algebra of $\sdiff S^2$ have been
computed in this basis in [9].
We note for future reference that the spherical harmonics can be
thought of as homogeneous polynomials in the variables $x_1$, $x_2$, $x_3$,
if $S^2$ is embedded in ${\bf R}^3$ by the equation $x_1^2+x_2^2+x_3^2=1$.
The three harmonics with $l=1$ (corresponding to the polynomials
$x_1$, $x_2$, $x_3$) generate the group $SO(3)$. In fact, defining
$$\eqalignno{
J_1=&-gx_1=-g\sin\theta \cos\varphi =
\ g\sqrt{2\pi\over3}(Y_1^1+Y_1^{-1})\ ,&(9a)\cr
J_2=&-gx_2=-g\sin\theta \sin\varphi =
           -ig\sqrt{2\pi\over3}(Y_1^1-Y_1^{-1})\ ,&(9b)\cr
J_3=&-gx_3=-g\cos\theta \phantom{\cos\varphi}=
-g\sqrt{4\pi\over3}Y_1^0 \ ,&(9c)\cr}$$
and using the Poisson bracket (4), we get
$\{J_\alpha,J_\beta\}=\varepsilon_{\alpha\beta\gamma}J_\gamma$.

For the quantization of a classical system with compact phase space
the best approach is that of geometric quantization [10]. This method
yields a Hilbert space and a set of quantum observables whose
algebra is isomorphic to the Poisson algebra of the corresponding
classical observables. Because of van Hove's theorem, this procedure
can not work for all observables. In the case of $S^2$ it works
successfully only for the functions $J_\alpha$.

Geometric quantization requires that $2g$ be an integer, {\it i.e.}
that the Dirac condition be satisfied. Then, the Hilbert space
${\cal H}_g$ is the space of polynomials of order $\leq 2g$ in
the complex variable $z=e^{i\varphi}\cot(\theta/2)$, with inner
product [11]
$$\langle\Psi,\Phi\rangle=
{2g+1\over2\pi i}\int_{S^2}{dz\,d\bar z\over(1+z\bar z)^{2g+2}}
\Psi^\ast(z)\Phi(z)\ .\eqno(10)$$
The dimension of ${\cal H}_g$ is $2g+1$ and a basis is given by
the functions $1,z,z^2,\ldots, z^{2g}$.

The operators corresponding to $J_1$, $J_2$ and $J_3$ are
$$\eqalignno{\hat J_1=&\ {1\over2}(1-z^2)\partial_z+gz
\ ,&(11a)\cr
\hat J_2=&\ {i\over2}(1+z^2)\partial_z-igz\ ,&(11b) \cr
\hat J_3=&\ z\partial_z-g\ ,&(11c)\cr}$$
obeying $[\hat J_\alpha,\hat J_\beta]=
i\varepsilon_{\alpha\beta\gamma}\hat J_\gamma$.
Notice that the value of the Casimir operator $\hat J^2$ in
the representation (11) is $g(g+1)$. Therefore, the Hilbert
space ${\cal H}_g$ is the familiar representation space of
angular momentum spanned by kets $|j,m\rangle$, with $j=g$.
The function $z^n$ in ${\cal H}_g$ corresponds to the state
$|g,n-g\rangle$.

As we have mentioned, the remaining functions on the sphere cannot be
quantized in such a way that their classical Poisson algebra is
preserved. Nevertheless, as in the case of ${\bf R}^2$, one could
envisage a more general scheme.

It follows from previous remarks that a basis for all smooth functions
on the sphere is given by polynomials
in the three harmonics with $l=1$. In order to achieve a quantization
of the whole set of classical observables it is therefore sufficient
to quantize powers of $J$'s. We associate to each monomial
$f^{(n_1,n_2,n_3)}=(J_1)^{n_1}(J_2)^{n_2} (J_3)^{n_3}$
an operator $\hat f^{(n_1,n_2,n_3)}$, in which the factors $\hat
J_\alpha$ have been ordered according to some given prescription.
In this way the whole algebra $\Gamma$ of functions on $S^2$ is turned
under quantization into the universal enveloping algebra
${\cal U}(so(3))$. The quantum symmetry algebra consists of the
operators in ${\cal H}_g$ corresponding to functions with vanishing
integral. These form the algebra ${\cal U}(so(3))/{\bf R}$, where
${\bf R}$ are the constant multiples of unity.

Since the Hilbert space is finite dimensional, this representation
will have an infinite dimensional kernel. A classical theorem of
Burnside [12] says that factoring out this kernel we remain with the
finite dimensional algebra of linear transformations in ${\cal H}_g$.
Thus one can regard the quantum symmetry algebra to be $sl(2g+1,{\bf C})$
or $sl(2g+1,{\bf R})$, depending on whether the representation of
$so(3)$ is complex or real.

\medskip
$N=S^{(1,1)}=SO(1,2)/O(2)$
\smallskip
\noindent
The hyperboloid $S^{(1,1)}$ is the surface in ${\bf R}^3$ defined
by the equation $x_1^2+x_2^2-x_0^2=-1$. On this surface
we define coordinates $\chi$, $\varphi$ by $x_1=\sinh\chi\cos\varphi$,
$x_2=\sinh\chi\sin\varphi$ and $x_0=\cosh\chi$.
As electromagnetic field we take ${\cal A}_\chi=0$,
${\cal A}_\varphi=g\cosh\chi$ and
${\cal F}_{\chi\varphi}=g\sinh\chi$. The symmetry group
$S({\cal F})=\sdiff S^{(1,1)}$ consists of volume-preserving
diffeomorphisms of $S^{(1,1)}$. Its maximal finite dimensional
subgroup is $SO(1,2)$. As in the case of the sphere, a basis for
functions on the hyperboloid is given by the homogeneous polynomials
in the variables $x_1$, $x_2$, $x_0$. The polynomials of degree
$\geq 1$ clearly give rise (through eq.(3)) to all hamiltonian
vectorfields, and close under Poisson bracket. Therefore also in this
case the Poisson algebra of the Noether charges does not get a central
extension.

The maximal compact subalgebra $so(1,2)$ is generated by the functions
$$\eqalignno{K_1=&\ gx_1=g\sinh\chi \cos\varphi\ ,&(13a)\cr
K_2=&\ gx_2=g\sinh\chi \sin\varphi\ ,&(13b)\cr
K_0=&\ gx_0=g\cosh\chi \ ,&(13c)\cr}$$
obeying $\{K_1,K_2\}=-K_0$, $\{K_0,K_1\}=K_2$, $\{K_0,K_2\}=-K_1$.
This subalgebra can be quantized without modification.
It is convenient to define
the complex coordinate $z=\tanh(\chi/2) e^{i\varphi}$ ($|z|<1$).
For $g\geq1/2$ we define the Hilbert space ${\cal H}_g$ as the space
of holomorphic functions on the unit disk with inner product
$$\langle\Psi,\Phi\rangle=
{(2g-1)\over2\pi i}\int_{|z|<1}dz\,d\bar z(1-z\bar z)^{2g-2}
\Psi^\ast(z)\Phi(z)\ .\eqno(14)$$
The condition on $g$ is dictated by the requirement that the inner
product be well defined (similar conditions have been discussed in
[13]). For $g<1/2$ one can use other representations of $so(1,2)$[14],
but we will not discuss this here.
The operators corresponding to $K_1$, $K_2$ and $K_3$ are [14,11]
$$\eqalignno{\hat K_1=&\ {1\over2}(1+z^2)\partial_z+gz
\ ,&(15a)\cr
\hat K_2=&\ {i\over2}(1-z^2)\partial_z-igz\ ,&(15b) \cr
\hat K_3=&\ z\partial_z+g\ .&(15c)\cr}$$
The value of the Casimir operator $\hat K_1^2+\hat K_2^2-\hat K_0^2$
in this representation is $g(1-g)$.
For non-integer values of $2g$, the above representation of the
algebra $so(1,2)$ leads to multi-valued representations of the group
$SO(1,2)$ [14].

Proceeding as in the case of the sphere, we now associate to each monomial
$f^{(n_1,n_2,n_3)}=(K_1)^{n_1}(K_2)^{n_2} (K_3)^{n_3}$
the operator $\hat f^{(n_1,n_2,n_3)}$ with a given ordering of the
factors $\hat K_\alpha$.
Then the quantum algebra of the Noether charges will be
${\cal U}(so(1,2))/{\bf R}$. It is represented faithfully on
${\cal H}_g$.

\medskip
The examples we have considered suggest that the quantization of
Chern-Simons quantum mechanics will follow the same general pattern,
at least for a wide class of manifolds $N$.
The classical symmetry group of canonical transformations has a
maximal finite dimensional subgroup $G$.
One can build a Hilbert space ${\cal H}$ carrying a realization of this group;
thus the algebra of the corresponding Noether charges
can be quantized in accordance with eq.(6).
We assume that all the symmetry charges (all functions on $N$)
can be expanded in polynomials in the generators of $G$
(this is called the ``strong generating principle'' in [15]).
Then, choosing any ordering, all these charges can be realized
as quantum operators on ${\cal H}$.
However, their algebra will now be different from
the classical one: it is the universal enveloping algebra of the Lie
algebra of $G$. This realization of the universal enveloping algebra
might have a nontrivial kernel, so to obtain a faithfully realized
symmetry algebra one has to factor it out. If $N$ is compact,
the resulting quantum symmetry algebra is finite dimensional.

We conclude by observing that this way of quantizing a classical
dynamical system is not restricted to Chern-Simons quantum mechanics,
but may be used also in the presence of a kinetic term.
Our approach allows to quantize all functions on phase space,
{\it i.e.} to construct a quantum operator acting on Hilbert space
for each classical observable. The obstruction to quantization
given in van Hove's theorem is circumvented by relaxing the condition
that (6) holds for all $f$, $g$.
As a consequence the algebra of quantum observables is a deformation
of the classical Poisson algebra and only reduces to the latter in the
classical limit.
This is different from the usual point of view on quantization,
where one insists on (6) but does not seek to quantize all functions
on phase space. Typically, one tries only to quantize those functions
which generate the symmetries of the theory.
When a kinetic or potential term is present, the symmetry group
is finite dimensional and, as we have observed, the algebra of its
generators is quantized without deformation. It is only for
``topological'' theories with vanishing Hamiltonian that every
function on phase space is a symmetry generator, and the necessity of
quantizing all observables arises.

\vfill\eject

\vskip 1in
\centerline{\bf REFERENCES}
\vskip 0.5in
\item{1.} R. Floreanini, R. Percacci and E. Sezgin,
Nucl. Phys. {\bf B 322} (1989) 255
\smallskip
\item{2.} G. V. Dunne, R. Jackiw and C. Trugenberger,
Phys. Rev. {\bf D 41} (1990) 661
\smallskip
\item{3.} P.S. Howe and P.K. Townsend, Class. Quantum Grav. {\bf 7}
(1990) 395
\smallskip
\item{4.} L.D. Faddeev and R. Jackiw, Phys. Rev. Lett. {\bf 60}
(1988) 1692
\smallskip
\item{5.} R. Abraham and J. Marsden, {\it Foundations of Mechanics},
(Benjamin Cummings, Reading, 1978)
\smallskip
\item{6.} S. Sternberg and J.A. Wolf, Trans. Am. Math. Soc. {\bf 238}
(1978) 1
\smallskip
\item{7.} J.W.B. Hughes, J. Math. Phys. {\bf 22} (1981) 245
\smallskip
\item{8.} M. Bordemann, J. Hoppe and P. Schaller, Phys. Lett. {\bf B
232} (1989) 199;\hfil\break
E. Bergshoeff, M.P. Blencowe and K.S. Stelle, Comm. Math.
Phys. {\bf 128} (1990) 213
\smallskip
\item{9.} J. Hoppe, Ph.D. thesis, MIT (1982);\hfil\break
T.A. Arakelyan and G.K. Savvidy, Phys. Lett. {\bf B 223} (1989) 41
\smallskip
\item{10.} Woodhouse, {\it Geometric Quantization}, (Cambridge
University Press, Cambridge, 1987)
\smallskip
\item{11.} A. Perelomov, {\it Generalized Coherent States and Their
Applications}, (Springer-Verlag, Berlin, 1986)
\smallskip
\item{12.} N. Jacobson, {\it Lectures in Abstract Algebra} II,
(van Nostrand, Princeton, 1953)
\smallskip
\item{13.} F.A. Berezin, Math. USSR Izv. {\bf 8} (1974) 1109; {\it ibid.}
{\bf 9} (1975) 341
\smallskip
\item{14.} V. Bargmann, Ann. Math. {\bf 48} (1947) 568
\smallskip
\item{15.} C.J. Isham, in {\it General Relativity, Groups and Topology II},
B. DeWitt and R. Stora, eds., (North-Holland, Amsterdam, 1984)
\vfill
\bye